\documentclass[twocolumn,preprintnumbers,amsmath,amssymb]{revtex4}

\usepackage{graphicx}
\usepackage{dcolumn}
\usepackage{bm}

\begin{document}

\preprint{APS/123-QED}

\title{  Detection of antiferromagnetic ordering in heavily doped LaFeAsO$_{1-x}$H$_x$ pnictide superconductors using nuclear-magnetic-resonance techniques
 }

\author{ N. Fujiwara$^1$\footnote {Email:naoki@fujiwara.h.kyoto-u.ac.jp}, S. Tsutsumi$^{1}$, S. Iimura$^{2, 3}$, S. Matsuishi$^{2, 3}$, H. Hosono$^{2, 3}$, Y. Yamakawa$^4$ and H. Kontani$^4$ }

\affiliation{$^1$Graduate School of Human and Environmental
Studies, Kyoto University, Yoshida-Nihonmatsu-cyo, Sakyo-ku, Kyoto
606-8501, Japan}

\affiliation {$^2$Material and structures laboratory (MSL), Tokyo
Institute of Technology, 4259 Nagatsuda, Midori-ku, Yokohama
226-8503, Japan }

\affiliation {$^3$Frontier Research Center (FRC), Tokyo
Institute of Technology, 4259 Nagatsuda, Midori-ku, Yokohama
226-8503, Japan}

\affiliation {$^4$Department of Physics, Nagoya University and JST, TRIP, Furo-cho, Nagoya 464-8602, Japan}





\begin{abstract}

We studied double superconducting (SC) domes in LaFeAsO$_{1-x}$H$_x$ by using $^{75}$As and $^{1}$H nuclear-magnetic-resonance techniques, and unexpectedly discovered that a new antiferromagnetic (AF) phase follows the double SC domes on further H doping, forming a symmetric alignment of AF and SC phases in the electronic phase diagram. We demonstrated that the new AF ordering originates from the nesting between electron pockets, unlike the nesting between electron and hole pockets as seen in the majority of undoped pnictides. The new AF ordering is derived from the features common to high-$T_c$ pnictides; however, it has not been reported so far for other high-$T_c$ pnictides because of their poor electron doping capability.

\end{abstract}

\maketitle

As a common feature in strongly correlated electron systems including high-transition-temperature (high-$T_c$) superconductors, the superconducting (SC) phase emerges adjacent to the antiferromagnetic (AF) phase on carrier doping. The AF order parameter vanishes by carrier doping, and the SC phase commonly manifests as a dome in the electronic phase diagram. On further carrier doping, the normal metallic state approaches a Fermi liquid state. This scenario is also seen in high-$T_c$ cuprates and several iron-based pnictides. In fact, a very close analogy between cuprates and pnictides such as  Ba(Fe$_{1-x}$Co$_x$)$_2$As$_2$ (Ba122 series)[1-3] was suggested in the electronic phase diagram[4]. The AF state appearing in an undoped or lightly doped regime is a stripe-type spin-density-wave (SDW) state[5] arising from interband nesting between hole and electron pockets, which is indicative of a superconducting mechanism via AF fluctuations[6, 7]. The nesting mechanism is supported by various experiments in the Ba122 series, in which an optimal doping level is located at the AF phase boundary, and therefore AF fluctuations are predominant even in the SC phase. However, LaFeAsO$_{1-x}$F$_{x}$ (La1111 series)[8], a prototype of iron-based high-$T_c$ pnictides, has an optimal doping level away from the AF boundary and the SC dome is separated from the AF phase boundary, as shown in Fig. 1(a)[8, 9]. Relatively high $T_c$ ($\sim$ 20 K) is maintained even in a F-overdoped regime ($x \thicksim 0.2$). These features would lend credence to an electron-phonon mechanism via orbital fluctuations[10, 11], another superconducting mechanism that is in debate with the AF fluctuation mechanism.

The unique features of LaFeAsO$_{1-x}$F$_{x}$ are highlighted by the newly-synthesized 1111 compounds CeFeAsO$_{1-x}$H$_{x}$ (Ref. [12]) and LaFeAsO$_{1-x}$H$_{x}$ (Ref. [13]). In these compounds, protons act as H$^{-}$ as well as F$^{-}$, and  O$^{2-}$ can be replaced with H$^-$ up to $x \sim 0.6$.  In the H-overdoped regime ($x \geq 0.2$), LaFeAsO$_{1-x}$H$_{x}$ undergoes a second superconducting state as shown in Fig. 1(a). Intriguingly, the first SC dome is very similar between F and H doping, suggesting that H doping supplies the same amount of electrons as F doping. The robustness of the second dome was confirmed by ac susceptibility measurements. The SC volume fractions for 30\% and 40\% H-doped samples are almost $100\%$.

The question is why the double SC domes manifest in this compound and what is the origin of the second SC dome. The primary purpose of this study was to investigate the second SC dome on a microscopic level by using the nuclear-magnetic-resonance (NMR) technique, but we unexpectedly discovered that a new AF phase follows the second SC dome on further H doping, as described below. We investigate the anomalous phase diagram having the double SC domes and two AF phases from both experimental and theoretical viewpoints.

The H-overdoped samples of LaFeAsO$_{1-x}$H$_{x}$ ($x=$0.53, 0.58 and 0.625) were prepared for the present measurements. The H doping level was analyzed by thermal desorption spectroscopy. The doping level observed by thermal desorption spectroscopy was almost the same with the nominal one for $x\leq0.30$, whereas the deviation between them ($\Delta x$) became large with increasing $x$. For $x\geq0.50$, $\Delta x$ reached $\sim0.1$. The powder samples were analyzed by x-ray diffraction. No crystallographical anomalies were observed for each doping level and the lattice constants changed monotonously with increasing $x$. The SC volume fractions of the 53\%, 58\% and 62.5\% doped samples were less than 12\%, 3\%, and 16\%, respectively. In the phase diagram, $T_c$s are plotted against the nominal doping level for the samples with the volume fraction of more than 20\%. Frequency detuning of the NMR tank circuit reflects the volume fractions. The resonance frequency of the NMR tank circuit shifts owing to the large demagnetization when cooled from the metallic state to the SC state. The shifts are plotted as percentages in Figs. 1(b)$-$1(d) for the 20\%, 40\%, and 62.5\% H-doped samples, respectively. One can find the largest shift for the 40\% H-doped samples which mark the optimal $T_c$.

\begin{figure}
\includegraphics{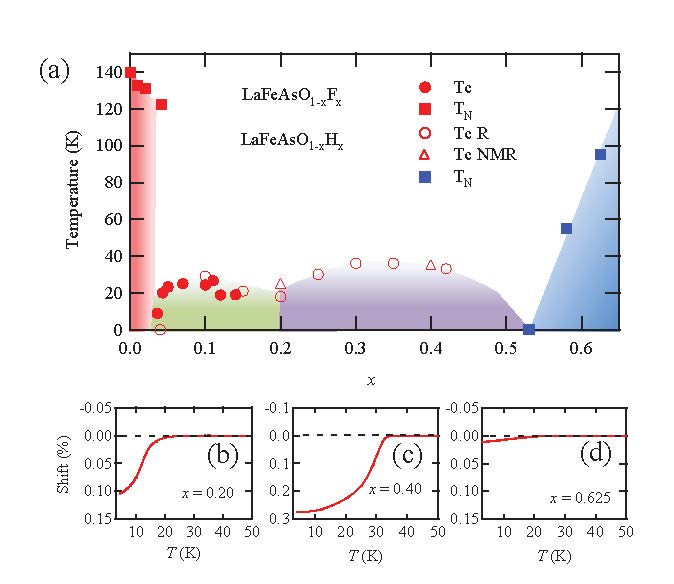}
\caption{\label{fig:epsart}  (color online) Phase diagrams of LaFeAsO$_{1-x}$F$_x$ and LaFeAsO$_{1-x}$H$_x$. (a)
The AF ordering temperatures ($T_N$) of LaFeAsO$_{1-x}$F$_x$ (closed red squares) were reported in magnetic-resonance measurements[5, 9]. The superconducting transition temperatures ($T_c$) of LaFeAsO$_{1-x}$F$_x$ and LaFeAsO$_{1-x}$H$_x$  were determined from resistivity measurements (closed and open circles)[8, 13] and the detuning of the NMR tank circuit (triangles). The AF ordering temperature for the heavily H-doped regime (closed blue squares) was determined from the linewidth of the $^1$H NMR spectra [Fig. 3(c)]. (b$-$d), Frequency detuning of the NMR tank circuit. Resonance frequency shifts when cooled from the metallic state. The bending points correspond to $T_c$.   }
\end{figure}

 The $^{75}$As NMR ($I$ = 3/2) spectra for the powder samples are illustrated in Fig. 2(a). The satellite signals ($I^z = -$3/2 $\Leftrightarrow$ $-$1/2 and $I^z$ = 1/2 $\Leftrightarrow$ 3/2)  spread in a wide region of the applied field ($\textbf{\emph{H}}$) depending on the angle between $\textbf{\emph{H}}$ and the principal axis of the electric field gradient (EFG) $V_{zz}$ at $^{75}$As sites[14]. The $V_{zz}$ direction is perpendicular to FeAs planes for iron-based pnictides, and the value is proportional to the nuclear quadrupole frequency $\nu_Q$.
 The central signal ($I^z$ = $-$1/2 $\Leftrightarrow$ 1/2) splits in two edges owing to a large $\nu_Q$ ($\sim 10-11$ MHz). The resonance signals should be symmetric with respect to the central signals, and thus the satellite signals ($I^z$ = 1/2 $\Leftrightarrow$ 3/2) should be observable at high fields. Unfortunately, the satellite signals overlap the $^{199}$La signals ($I$ =5/2) at the NMR frequency of 35.1 MHz, as shown in Fig. 2(b). In an ideal case, a central peak ($I^z$ = $-\frac{1}{2}\Leftrightarrow \frac{1}{2} $) and four satellite edges ( $I^z$ = $\pm\frac{1}{2}\Leftrightarrow \pm\frac{3}{2} $ and $\pm\frac{3}{2}\Leftrightarrow \pm\frac{5}{2} $) should be observable for $^{199}$La owing to quadrupole interaction; however, some distribution of EFG makes the edges unclear and the line shape looks like a single broad peak.  The relative intensity of the $^{199}$La signals is apparently larger than that of the $^{75}$As signals because the former signals are distributed in a narrow $H$ region owing to small quadrupole interaction.

 \begin{figure}
\includegraphics{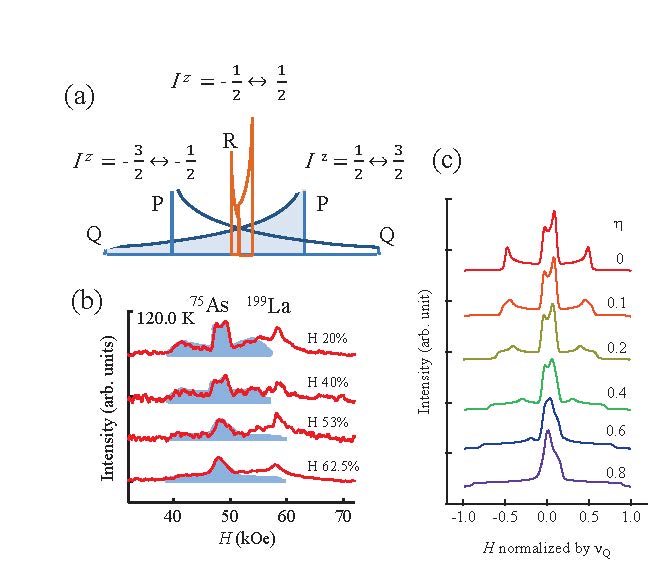}
\caption{\label{fig:epsart} (color online) Orbital ordering or differentiation of $d$ orbital weight at each Fe site (a) Illustration of NMR powder pattern ($\eta = 0$) for $I=\frac{3}{2}$ under a fixed frequency.  The signals corresponding to the transition $I^z=\frac{1}{2}\Leftrightarrow\frac{3}{2}$ are shaded in blue. The edges $P$ and $R$ come from the powder samples with FeAs planes parallel to the applied field.  The edge $Q$ comes from the samples with FeAs planes perpendicular to the applied field. (b) NMR spectra measured at 35.1 MHz. $^{75}$As signals are shaded in blue. (c) Simulation of the line shape for $I=\frac{3}{2}$. The horizontal axis is normalized by $2\pi\nu_Q/ ^{75}\gamma_N$, where $^{75}\gamma_N$ (= 7.292 MHz/10 kOe) is the gyromagnetic ratio of $^{75}$As.  }
\end{figure}

Interestingly, the anisotropy of EFG defined by

\begin{equation}
\eta  \equiv \frac{V_{yy}-V_{xx}}{V_{zz}},
\end {equation}

where $V_{yy}$ and $V_{xx}$ represent EFGs in FeAs planes, increases with increasing $x$. On H doping, the two edges at 48-50kOe become an asymmetric peak, and the satellite signal at 41-42kOe becomes unclear. The evolution is caused by the increase in $\eta$ as seen from the powder-pattern simulations in Fig. 2(c). We followed the procedure in Ref. 15 to simulate the powder patterns. The values of $\eta$ are estimated to be 0.1-0.2 for the 20\%-40\% H-doped samples, whereas they are $\sim$0.6 for the 53\%-62.5\% H-doped samples. The increase in $\eta$ is characteristic of the H-overdoped regime; for LaFeAsO$_{1-x}$F$_{x}$, $\nu_Q$, namely, $V_{zz}$ increases linearly on F doping, and $\eta$ remains almost unchanged. The EFGs are determined from the surrounding ions and $d$ electrons in FeAs planes: the contribution of the former remains unchanged on H doping considering the lattice parameters, and thus the increase in $\eta$ is attributable to the in-plane anisotropy coming from the differentiation of the orbital weight among the $d_{xy}$, $d_{yz}$, and $d_{zx}$ orbitals at each Fe site. We will discuss the possibility of orbital ordering later.

\begin{figure}
\includegraphics{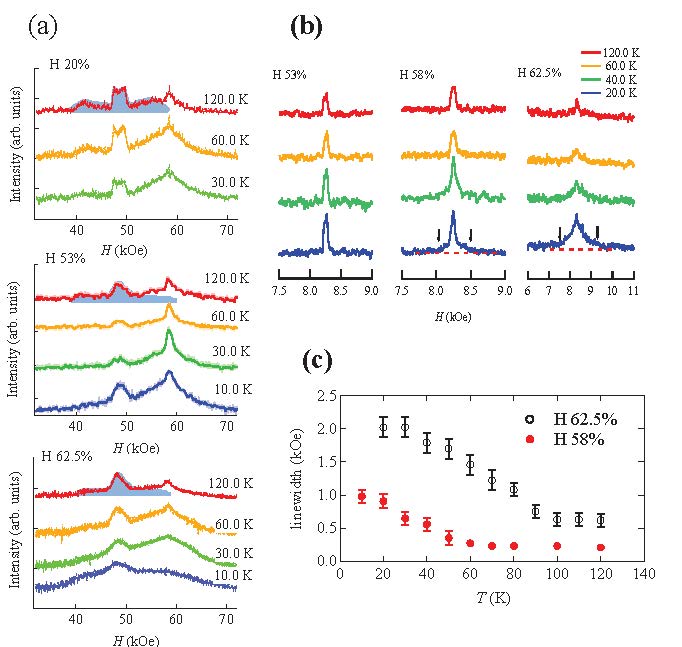}
\caption{\label{fig:epsart} (color online) AF ordering observed for 58\% and 62.5\% H-doped samples.(a) $^{75}$As NMR spectra at several temperatures.  (b) $^{1}$H NMR spectra at several temperatures. (c) Temperature dependence of the linewidth defined by the $H$ span between arrows in Fig. 3(b). }
\end{figure}

\begin{figure}
\includegraphics{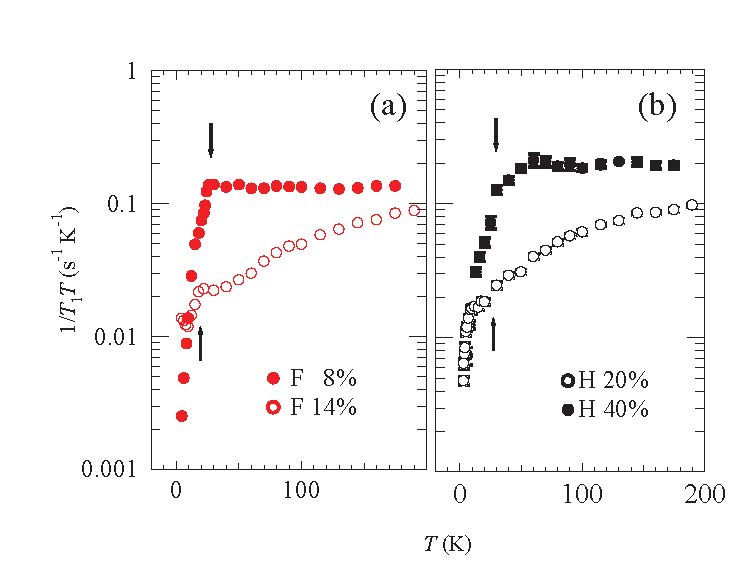}
\caption{\label{fig:wide} (color online) Nuclear magnetic relaxation divided by temperature $1/T_1T$ of $^{75}$As. $1/T_1T$ gives a measure of spin fluctuations.  (a) $1/T_1T$ of $^{75}$As of LaFeAsO$_{1-x}$F$_x$. (b) $1/T_1T$ of $^{75}$As of LaFeAsO$_{1-x}$H$_x$.  Arrows indicate $T_c$.  }
\end{figure}

Another remarkable feature of the heavily H-doped regime ($x \geq 0.58$) is the revival of some AF ordering, as shown in Fig. 3(a). The broadening of the line shape is not observed for the 20\% and 53\% H-doped samples, whereas it is remarkable at low temperatures for the 62.5\% H-doped samples. We confirmed the broadening from $^1$H NMR ($I$ = 1/2) spectra. The $^1$H signals are free from nuclear quadrupole interaction owing to $I$ = 1/2, and thus are sensitive only to AF ordering. Figure 3(b) shows $^1$H NMR spectra for the H-overdoped samples. The 62.5\% H-doped samples exhibit considerable broadening at low temperatures even though the linewidth for the 53\% H-doped samples is almost unchanged. O$^{2-}$ can be replaced only with H$^-$ in LaFeAsO, and therefore the broadening of the 62.5\% H-doped samples is intrinsic and demonstrates the emergence of some AF ordering. The ordering temperature is determined from the temperature dependence of the linewidth, as shown in Fig. 3(c). Unlike the AF ordering in the lightly doped regime, the AF ordering in the heavily H-doped regime is not spatially uniform. If the ordered moments were spatially uniform, the line shape would be of the rectangular type [16]. We calculated the internal field at a proton nucleus for the nearest neighboring 42 iron sites assuming the stripe-type spin configuration in the neighboring FeAs planes and assuming the same spin alignment to the direction normal to the FeAs planes. From the internal fields indicated by arrows in Fig. 3(b), the maximal spin moments are estimated to be 1.3$\mu_B$ and 2.4$\mu_B$ for the 58\% and 62.5\% H-doped samples, respectively. These values are remarkably larger than those in the first AF state (0.36$\mu_B$) [5]. The spin moments in the second AF state are also larger than those for other pnictides (0.3-0.9$\mu_B$) [17-19] and are comparable to those for K$_x$Fe$_{2-y}$Se$_2$ ($\sim$2.8$\mu_B$)[20]. The emergence of the second AF ordering is very anomalous for strongly correlated electron systems because, in general, carrier doping breaks AF ordering.


Next, we investigated the AF ordering from the relaxation rate per temperature $1/T_1T$. The quantity gives a measure of AF fluctuations and is enhanced under strong AF fluctuations. Figures 4(a) and 4(b) show $1/T_1T$ of $^{75}$As for the F-underdoped and H-overdoped regimes, respectively, measured at the edge $R$ in Fig. 2(a). For the 20\% H-doped samples, $1/T_1T$ is almost the same as that for the 14\% F-doped samples at high temperatures above $T_c$. On H doping up to 40\%,  $1/T_1T$ is enhanced remarkably and shows almost the same $T$ dependence as that for the F-underdoped regime (see the 8\% F- doped samples)[21-24], indicating that AF fluctuations revive with the approach of the second AF phase on further H doping. The results are consistent with the broadening of the NMR spectra for the heavily H-doped samples.

\begin{figure*}
\includegraphics{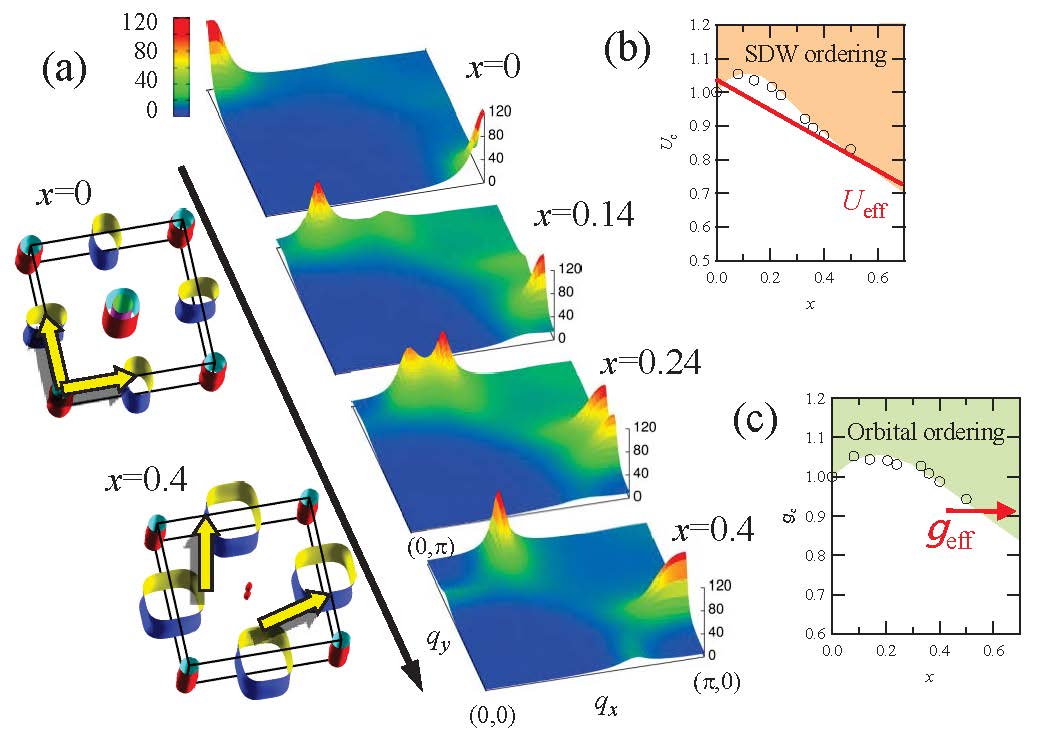}
\caption{\label{fig:epsart} (color online) Theoretical investigations for LaFeAsO$_{1-x}$H$_x$. (a) Fermi surfaces and 2D maps of the spin susceptibility calculated by the tight binding model under RPA. Coulomb interaction couplings used for the calculations are a few percent smaller than critical Coulomb interaction couplings ($U_c$). The nesting between Fermi surfaces shown by the arrows makes peaks on the 2D maps, causing the first and second AF orderings. (b) Doping dependence of $U_c$. The value of $U_c$ for each doping level is normalized by that for the end material ($x=0$). (c) Doping dependence of critical quadrupole interaction couplings ($g_c$ ). The value of $g_c$ is also normalized by that for the end material ($x=0$). The Coulomb and quadrupole interaction couplings ($U_{eff}$ and $g_{eff}$) are larger than the critical values for the heavily H-doped samples, leading to the AF and orbital orderings.
  }
\end{figure*}

The present NMR results revealed the unusual AF ordering. To understand the phenomenon, we calculated the spin susceptibility based on the tight binding model under random phase approximation (RPA). For each doping level, AF ordering is potentially possible for the Coulomb interaction coupling ($U$) stronger than the critical coupling $U_c$. Figure 5(a) shows the two-dimensional (2D) color maps of the spin susceptibility calculated for $U$ a few percent smaller than $U_c$. For the undoped samples, the nesting between electron and hole pockets, as shown by the arrow in the illustration of Fermi surfaces for $x=0$, makes a peak at ($\pi$, 0) on the 2D map, causing the first AF ordering. The ($\pi$, 0) nesting weakens and the other nesting between electron pockets shown by the arrow in the illustration for $x=0.4$ develops on H doping.  The peak position deviates from  ($\pi$, 0), suggesting an incommensurate SDW state. The incommensurate state is consistent with the cusp-type $^1$H-NMR spectra. As shown in Fig. 5(b), the doping dependence of $U_c$ exhibits an upward convex curve due to the change in the nesting condition, leading to the reappearance of AF ordering at $x>0.25$ if the effective coupling $U_{eff}$ is constant. In fact, $U_{eff}$ would moderately decrease with increasing $x$,  because the Kanamori theory predicts $U_{eff}(x) \sim U/(1+UN(x))$[25], where $N(x)$ is the density of states at the Fermi level, and it slightly increases with increasing $x$ in the band calculations$[13]$. $U_{eff}(x)$ can be approximated by a linear function of $x$ under the condition $UN(x) \ll 1$. As shown in Fig. 5, one can draw the line of $U_{eff}$ which crosses the $U_c$ curve at two AF quantum critical points, and thus Figure 1a is reproduced. In Fig.5, we used two experimentally-determined quantum critical points, $x\sim 0.04$ and $\sim 0.58$, to draw the line.

The anisotropy $\eta$ is enhanced on H doping prior to the second AF ordering. According to the band calculation[13], the partial density of states associated with the $d_{xz, yz}$ and $d_{xy}$ orbitals is almost unchanged for $0.4\leq x \leq0.625$. Furthermore, mixing of the $d$ orbitals tends to make $\eta$ small. Therefore, the enhancement of $\eta$ on H doping [Fig. 2(b)] suggests spontaneous alignment of the $d$ orbitals at each Fe site, being indicative of some orbital ordering. Strong orbital fluctuations can be induced by the charge quadrupole interaction coupling $g$ originating from the many-body effect by $U$ beyond the mean-field level in addition to the electron-phonon interaction.The orbital ordering is realized when the effective coupling $g_{eff}$ is larger than the critical coupling $g_c$. The doping dependence of $g_c$ is also a convex curve, as shown in Fig. 5(c), and therefore orbital ordering tends to manifest in a heavily H-doped regime similarly to the AF ordering. The value of $g_{eff}$ is estimated to be $\sim$0.9 for the heavily H-doped regime because the orbital ordering is expected to occur for $x\geq0.53$ in the present experiments.

The doping dependence of $g_c$ and $U_c$ is upward convex and both quantities become small, in other words, spin and/or orbital fluctuations are enhanced in the lightly and heavily H-doped regimes; this gives the answer for why the double SC domes manifest between the lightly and heavily H-doped regimes. Despite the fact that the observed phenomena are derived from the features common to high-$T_c$ pnictides, these phenomena have not been observed for other high-$T_c$ pnictides. Our findings are due to the capability of electron doping in LaFeAsO$_{1-x}$H$_{x}$.

In conclusion, we discovered that the new AF ordering emerges adjacent to the second SC dome on heavy H doping, forming a symmetrical alignment of the AF and SC phases in the electronic phase diagram, as shown in Fig. 1(a). From the calculations based on the tight binding model within RPA, we demonstrated that the nesting between electron pockets can induce the AF ordering in a heavily electron doped regime. We also observed the enhancement of $\eta$ prior to the emergence of the new AF ordering. The enhancement suggests differentiation of $d$ orbitals weight or some orbital ordering. The answer for why the double SC domes manifest is attributed to the enhancement of spin and/or orbital fluctuations in lightly and heavily H-doped regimes.

The NMR work is supported by a Grant-in-Aid (Grant No. KAKENHI 23340101) from the Ministry of Education, Science, and Culture, Japan. This work was supported in part by the JPSJ First Program. We thank T. Morinari for discussion, and R. Sakurai for experimental assistance.







\end{document}